# MigrantStore: Leveraging Virtual Memory in DRAM-PCM Memory Architecture

Hamza Bin Sohail, Balajee Vamanan and T. N. Vijaykumar
School of Electrical and Computer Engineering, Purdue University
{hsohail,bvamanan,vijay}@ecn.purdue.edu

## Abstract

With the imminent slowing down of DRAM scaling, Phase Change Memory (PCM) is emerging as a lead alternative for main memory technology. While PCM achieves low energy due to various technology-specific advantages, PCM is significantly slower than DRAM (especially for writes) and can endure far fewer writes before wearing out. Previous work has proposed to use a large, DRAM-based hardware cache to absorb writes and provide faster access. However, due to ineffectual caching where blocks are evicted before sufficient number of accesses, hardware caches incur significant overheads in energy and bandwidth, two key but scarce resources in modern multicores. Because using hardware for detecting and removing such ineffectual caching would incur additional hardware cost and complexity, we leverage the OS virtual memory support for this purpose. We propose a DRAM-PCM hybrid memory architecture where the OS migrates pages on demand from the PCM to DRAM. We call the DRAM part of our memory as *MigrantStore* which includes two ideas. First, to reduce the energy, bandwidth, and wear overhead of ineffectual migrations, we propose *migration hysteresis*. Second, to reduce the software overhead of good replacement policies, we propose *recently-accessed-page-id (RAPid) buffer*, a hardware buffer to track the addresses of recently-accessed MigrantStore pages.

## 1 Introduction

Since the advent of VLSI, DRAM technology has continued to scale in density and cost even faster than Moore'a Law to provide larger and cheaper main memory in modern computer systems. However, experts agree that DRAM's charge-based approach to providing storage is not likely to scale beyond a few more technology generations (e.g., 20nm) [14]. The end of DRAM scaling has prompted researchers to investigate a number of alternative storage technologies such as spin torque transfer RAM (STTRAM), Flash, and phase-change memory (PCM). To store a binary value, STTRAM, Flash, and PCM, respectively, use spin, electrically-isolated charge, and phase (crystalline versus amorphous). Because these properties do not depend on electric power to be retained, these storage technologies are non-volatile. Among the alternatives, PCM is emerging as the lead contender due to its superior combination of energy, speed, density, and reliability (wear) characteristics. PCM is about two orders-of-magnitude faster and can endure about four orders-of-magnitude more writes than Flash, and dissipates two orders-of-magnitude less energy and is two-orders-of-magnitude denser than STTRAM.

While PCM is projected to scale in density well beyond the technology nodes where DRAM scaling stops, PCM is not without challenges. Compared to DRAM, PCM is about four times slower for reads and about twelve times slower for writes; and can endure only about $10^9$ - $10^{12}$ writes before wearing out [14]. These longer latencies also imply longer bank busy times and hence lower bandwidth. Though PCM is likely to improve significantly before DRAM completely stops scaling, PCM's reliability and performance (latency and bandwidth) will likely continue to require improvements via architectural and system support. PCM dissipates about forty-times more energy per cell for writes and about twice as much energy per cell for reads. Fortunately, PCM's technology offers other advantages over DRAM, such as significantly less leakage, non-destructive reads, and writes that can selectively update a part of a row in a memory array without requiring additional selection circuitry within the array which would degrade density. These advantages more than offset the per-cell energy disadvantage enabling PCM to achieve energy comparable to DRAM. Recent architecture papers [7][12][19] have proposed several hardware schemes to exploit PCM's advantages (scalability and energy) and alleviate its disadvantages (performance and reliability). While these papers propose interesting ideas for PCM-based memory systems, we take a different approach in this paper.

Qureshi et al. [12] propose to improve performance and wear by employing a large, DRAM cache in hardware (e.g., 128 MB). However, the bandwidth demand on and the energy dissipation of the PCM are worsened if the DRAM-cached blocks are not accessed enough. We make the key observation that because of good caching at the L1 and L2, many blocks are evicted from the DRAM cache before being accessed enough to amortize the bandwidth and energy cost of caching the block in the DRAM cache. Further, most of the data is written only once before eviction from the DRAM cache so that the PCM is written only once with or without DRAM caching and hence such *ineffectual* cache fills do not reduce PCM wear. In fact, by evicting useful, dirty pages from the DRAM cache, the ineffectual cache fills increase writebacks to the PCM, worsening PCM wear. Bandwidth and energy are two key but scarce resources in modern multicores. To make matters worse, PCM's long latencies imply long bank occupancies which degrade bandwidth despite aggressive banking; Unfortunately, multicores can absorb latencies via thread overlap but cannot compensate for lack of bandwidth.

For performance, energy, and wear, large buffering using DRAM appears to be a reasonable approach. However, we employ intelligent OS policies to avoid the hardware cache's ineffectual cache fills. We propose a DRAM-PCM hybrid memory architecture that leverages virtual memory so that the OS migrates pages on demand from the PCM to DRAM. We call the DRAM part of our memory as *MigrantStore*. Being a part of physical memory, MigrantStore's access is handled by virtual address translation. The DRAM in [12] is a hardware cache and not part of OS-managed physical memory.

Conventional OS placement and replacement policies for managing physical memory cause MigrantStore to incur significant performance and energy degradations. We propose two ideas to address these issues. First, like the hardware cache, MigrantStore also incurs the bandwidth and energy cost of ineffectual migrations. The software overhead of the migrations further degrade system performance and energy. To address this issue, we propose to isolate the heavily-accessed pages via our *migration hysteresis* where a page is placed (i.e., migrated) only after some number of accesses have occurred to the page in PCM. One may think that the hysteresis can be implemented in the hardware cache as well. However, doing so would require tracking pages that are not in the DRAM cache but in the PCM main memory. Because potentially a large fraction of the main memory may need to be tracked, implementing the hysteresis in the harware cache would be impractical whereas the OS is a natural choice for such tracking.

Second, current OS's periodically scan the reference bits in the page tables to identify recently-accessed pages for implementing replacement policies for main memory. However, being a small fraction of physical memory, MigrantStore incurs much more frequent replacements than typical main memory. (Because of DRAM's worse scaling than PCM, MigrantStore would always be a small fraction of physical memory.) To keep the OS scan overhead low, the scans are done at coarse time granularities (e.g., every 30 seconds), whereas the migrations occur more frequently (e.g., every 10-20 microseconds at 2 GHz). While the relatively-infrequent scanning would render the replacement information stale, using random replacement would incur significant performance loss. To address this issue, we propose a small hardware buffer, called *recently-accessed-page-id (RAPid) buffer*, to hold (a truncated set of) the addresses of the MigrantStore pages accessed between two migrations so that the replacement software can scan the buffer instead of the page tables. We exploit the fact that because migrations are frequent, *RAPid buffer* can be small (e.g., 20 entries).

The key contributions of this paper are:
- *MigrantStore*, a DRAM-PCM hybrid memory architecture that leverages virtual memory to reduce cost and energy overhead compared to hardware caching;
- *migration hysteresis*, a placement policy, to reduce the bandwidth, and energy overhead of ineffectual migrations;
- *recently-accessed-page-id (RAPid) buffer* to reduce the software overhead of good replacement policies; and
- Using simulations of commercial workloads, we show that MigrantStore performs better than all the compared alternative schemes and consumes less energy than the DRAM-based schemes. For instance, a 128-MB MigrantStore improves energy and performance by 25% and 7%, respectively, over a 128-MB hardware cache. while achieving similar wear.

The rest of the paper is organized as follows. We provide background on PCM and discuss related work in Section 2. We describe MigrantStore in Section 3. We explain our methodology in Section 4. We show our results in Section 5 and conclude in Section 6.

## 2 Phase Change Memory (PCM): Background and Related Work

While DRAM is a charge-based storage technology, PCM uses phase — crystalline or amorphous — to store state. The two phases exhibit different resistivities that can be detected to determine the phase. PCM retains its phase even in the absence of electric power, making PCM a non-volatile storage medium. There are diode- [6] and transistor-based [2] PCM implementations. Because the supply voltage scales better in transistor-based implementations [6], we assume transistor-based implementation in the rest of the paper.

Each PCM cell is made of a BJT transistor and a storage element (a resistor) between the BJT's emitter and the bitline [3] (see Figure 1). The BJT's base is connected to the wordline and the collector is connected to ground. Reads cause current flow through the storage element due to a voltage applied at the bitline. This current, which varies depending on the storage element's state, is sensed at the end of the bitline to determine the cell's state. For writes, instead of sensing the current flow through the storage element, a current is sent through the bitline to heat the storage element, causing the element to change its phase. The magnitude and duration of the current flow determines the element's resultant phase. We now discuss PCM's performance, energy, and wear.

### 2.1 Performance

A DRAM or PCM access includes not only the cells but also address decoders, row buffers, and selection out of the row buffer. These components are common between DRAM and PCM so that the differences in the *total* access latency and energy between DRAM and PCM are less than the differences in the *per-cell* latency and energy. While PCM per-cell read and write latencies are, respectively, about four and twelve times longer than DRAM per-cell latencies [2][6][3][17][4], the total PCM read and write latencies are, respectively, only about two and six times as slow as DRAM.

Writes to PCM occur in the background upon writebacks from the on-chip caches, off the program critical path (except in the uncommon case of filling up of writeback buffers). Consequently, PCM's long write latencies may not impact performance by much. While reads are critical, the 2x read latency does not translate to doubling of read miss penalty which includes latencies and queuing delays at the memory controller and the memory bus in addition to the raw read latency. Typically, the raw latency contributes only 33-50% of the miss penalty, and hence the miss penalty may go up by 1.16-1.5x.

The longer read and write latencies fundamentally imply that PCM banks remain busy for longer times. This longer occupancy reduces bandwidth even in heavily-banked PCM memories where bank conflicts are inevitable. While PCM's longer latencies can be hidden by multi-threading, such hiding exacerbates bandwidth pressure which is a significant concern.

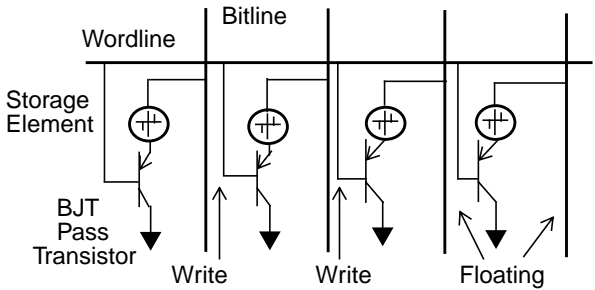

**FIGURE 1: PCM cell and selective update**

### 2.2 Energy

Like latencies, the *total* energy cost of a PCM access relative to a DRAM access is less than the *per-cell* energy cost of a PCM access relative to that of a DRAM access. Because PCM reads are non-destructive, unlike DRAM reads, the read row need not be written back. Therefore, though PCM per-cell read energy is about twice as much as DRAM per-cell energy [2][6][3][17][4], the total PCM read energy is about as much as that of DRAM. PCM writes, however, need to heat the storage element to change its phase, in contrast to DRAM writes which change the charge on a storage capacitor. The current flow for writes is much higher than that for reads — PCM per-cell write energy is about forty times as much as that of DRAM [3] (diode-based implementations may incur less write energy but they do not scale well, as mentioned above). However, the total PCM write energy is about twenty times as much as that of DRAM. Further, PCM's ability to update selectively parts of a row more than offsets this factor of twenty.

Fundamentally, PCM's selective updates stems from the fact that PCM is phase-based whereas DRAM is charge-based. In a PCM write, though the wordline is fired causing all the BJTs in the entire row to be turned on, bitlines in parts of the row need not carry any current and can simply float without affecting the phase of the storage elements in those parts (see Figure 1). Consequently, selective update in PCM can occur simply in the write driver circuitry at the periphery of the array without changing the wordline circuitry. In contrast, DRAM is charge-based, and hence the DRAM cells in those parts of the row would simply discharge through the floating bitlines and would corrupt the stored state. Hence, selective update in DRAM (or SRAM) would require selective activation of the wordlines so that the parts of the wordline connected to the cells that are not written would not be fired. Such selective activation would require extra control wires and circuitry *within* the array, severely degrading density. Therefore, DRAM designers do not employ selective update and instead rely on sizing the arrays such that the array is small enough that reading and writing to the whole row does not require high energy and at the same time the array is large enough that the wiring overhead does not increase the area significantly. Due to selective updates, PCM nearly eliminates this area-energy trade-off by allowing the PCM designer to size the array based almost *solely* on area constraints while using selective update to tackle energy *independently* (reads can also be done selectively but read energy is much less than write energy).

As such, because most writes to main memory (DRAM or PCM) are much narrower than a row, selective update achieves a large reduction in the write energy (e.g., 64-byte L2 block versus 8-KB PCM row spanning the DIMM modules amounts to an 128x reduction).

Because PCM is not charge-based, PCM's leakage is significantly lower than that of DRAM (e.g., 10x lower) [13][2][6]. This advantage is substantial given that DRAMs incur significant leakage (e.g., more than 15% of the total energy comes from leakage in non-standby mode) [16]. Together with selective update, the lack of leakage enables PCM to be nearly equal to DRAM in total access energy despite being worse in the per-cell energies.

### 2.3 Wear

Because writes involve heating the storage element to change its phase, writes cause PCM cells to wear out. In general, PCM cells can endure about $10^9$ - $10^{12}$ writes whereas DRAM cells wear out after about $10^{16}$ writes [14]. If left unaddressed, PCM used in main memory could wear out in a few months depending on the applications' L2 writeback behavior. Selective updates help reduce the wear by avoiding unnecessary writes. In addition, large buffering, as provided by MigrantStore or the hardware cache in [12] can absorb a significant fraction of the writebacks.

### 2.4 Related Work

Starting with the pioneering work on PCM [17], there has been significant work on technology, devices, and materials for PCM [2][6][3][13][4]. There have been some architecture proposals to alleviate PCM's reliability and performance problems.

Zhou et al. [19] propose (1) fine-grain wear leveling by rotating the PCM row upon every write; (2) coarse-grain wear leveling by swapping segments after some number of writes; and (3) reducing write energy (and bit-level wear) by using bit-level comparisons to avoid writing the same bit values via selective update. However, the scheme requires a large barrel shifter per PCM chip for the row rotation (e.g., 4K bytes), in addition to hardware tables for maintaining (1) the rotation amounts per PCM row and (2) the mapping between segments' physical address to segments' physical location. For large memories, the hardware tables, which are as large as page tables, add significant overhead and complexity. In contrast, MigrantStore exploits existing address translation mechanisms without needing any extra hardware tables.

Lee et al. [7] propose to reduce PCM write energy and wear by optimizing PCM's row buffers so that multiple writes to the same location are absorbed in the buffers resulting in only a single writeback to the PCM array. The authors propose more and narrower buffers than DRAMs to balance temporal and spatial locality. However, PCM wears out so much sooner than DRAM (e.g., six orders of magnitude fewer writes) that a large fraction of PCM writes need to be absorbed requiring large buffers. Unfortunately, the row buffers take up area and cannot be increased beyond a few megabytes, and are inflexible in that a given row buffer can hold data only from its own subarray and not from other subarrays. We show that the row buffers *alone* are insufficient for commercial applications using multi-gigabyte PCM memory. In contrast, MigrantStore and the hardware cache [12] are more plentiful in capacity and flexible in mapping, and hence are more effective.

In addition to proposing the hardware cache, Qureshi et al. [12] observe that in their database workload, writes to the cache block at the top of a page (i.e., block #0 in each page) are more frequent than writes to the rest of the blocks. Accordingly, they propose leveling this uneven wear of cache blocks by rotating the blocks within a physical page by a random amount, at the time the page is swapped into memory from disk. The rotation amount is held in a hardware table in the memory controller. Unfortunately, the table is as large as virtual memory page table and imposes significant overhead. Despite many attempts, however, we did not observe any systematic bias in our commercial workloads. Consequently, we do not rotate the blocks within a page, and avoid this overhead.

While another work [18] proposes copying of frequently-written pages from PCM to DRAM at the end of OS quanta, the scheme does not leverage on-demand migration, a fundamental feature and cornerstone of modern virtual memory. As such, the scheme ends up copying pages accessed in a quantum at the end of the quantum by which time the pages are no longer frequently accessed while incurring PCM latencies when the pages are being accessed — i.e., OS quanta are too long during which locality changes significantly. Further, copying frequently-written pages ignore frequently-read pages which incur PCM latencies. We show that the scheme is not effective in improving performance or wear.

A few other papers address PCM's wear problem [11,15,5]. Another work targets PCM's slow writes by pausing writes to allow intervening reads [10]. While these four schemes do not reduce but either spread out wear or reschedule writes, the hardware cache and MigrantStore use DRAM to reduce the write traffic to the PCM improving not only wear but also performance and energy.

## 3 MigrantStore

To avoid the hardware cache's ineffectual cache fills, we employ intelligent OS policies for a DRAM-PCM hybrid memory architecture that leverages virtual memory. As described in the Section 1, the OS migrates pages on demand from the PCM to DRAM which is called the *MigrantStore* (see Figure 2). Due to locality, a small MigrantStore, when compared to PCM, would be effective (e.g., for an 8GB PCM, a 128-MB MigrantStore — less than 2% in size — may suffice). Because of MigrantStore's small size, the physical memory or the page tables do not grow much.

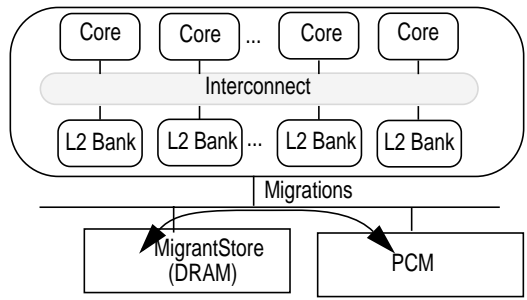

**FIGURE 2: MigrantStore**

Conventional OS policies for managing physical memory cause MigrantStore to incur significant performance and energy degradations. Recall from Section 1 that to address these issues, we propose two ideas: (1) *migration hysteresis* to reduce bandwidth and energy overhead of ineffectual migrations where the migrated pages are evicted before being accessed enough number of times in MigrantStore; (2) *recently-accessed-page-id (RAPid) buffer* to reduce the replacement policy software overhead of scanning page tables to identify recently-accessed MigrantStore pages. By triggering fewer migrations, the hysteresis also reduces the OS software overhead.

### 3.1 MigrantStore Operation

An access to a page in PCM triggers a fault causing the page to be migrated to MigrantStore (page faults directly map the page into MigrantStore). Because physical memory includes both PCM and MigrantStore, we need to identify whether a page is in PCM or MigrantStore. To this end, the OS sets a bit in the page table entry when a page is migrated or swapped into MigrantStore, and clears the bit when the page is evicted out of MigrantStore.

Unlike a page fault which must retrieve data from the slow disk, PCM faults transfer data between storage media that are much faster than the disk — from the PCM to MigrantStore. Accordingly, we assume a fast trap for PCM faults. The trap handler invokes a DMA to copy the page from PCM to MigrantStore, and to write back the evicted MigrantStore page (if dirty) to the PCM. Two points similar to today's systems are: (1) Because a page contains contiguous addresses, the DMA achieves high bandwidth via open-page mode and cache-block-size bursts to exploit row locality in both the PCM and MigrantStore's DRAM. (2) Stale cache blocks belonging to the copied pages are flushed from the caches. The DMA frees the CPU to track the replacement priority of MigrantStore pages using the RAPid buffer. Because both PCM and MigrantStore are reasonably fast, switching to another thread during a migration in addition to or instead of replacement-policy bookkeeping may incur higher overhead. Accordingly, our experiments assume that the processor does not switch threads and instead runs the replacement-policy software during migrations (Section 3.2). While the migrations inev-itably incur some performance and energy overhead, our migration hysteresis reduces this overhead for ineffectual migrations (Section 3.3).

MigrantStore can be located on the system memory bus along side PCM, or behind the L2 on a dedicated link similar to an L3 cache. Irrespective of its placement, MigrantStore does not raise any new coherence issues for I/O because MigrantStore is part of physical memory and pages in MigrantStore may be involved in I/O similar to PCM pages. Further, MigrantStore does not need any extra processor pins because it shares the processor pins with the PCM. For high bandwidth, MigrantStore is banked, just as the PCM is banked.

The key hardware cost of MigrantStore is the DRAM. As discussed before, because MigrantStore is much smaller than the PCM, the page table overhead is small. While the usual address translation locates the desired page in MigrantStore, the hardware cache in [12] relies on tags and incurs higher cost overhead.

### 3.2 Replacements

Recall from Section 1 that migrations occur too frequently to allow the MigrantStore replacement software to scan the reference bits in the page tables at every migration for avoiding replacement of recently-accessed MigrantStore pages. Unfortunately, random replacement instead of any of the stack-based policies evicts useful pages and incurs considerable performance and energy loss. To address this issue, we propose a hardware buffer, called the RAPid buffer, to hold the addresses of MigrantStore pages accessed between two consecutive migrations. The buffer is placed in the memory controller which inserts the page addresses in hardware. During the DMAs of each migration, the replacement module in the OS scans the RAPid buffer instead of the full page tables to update its data structures (e.g., LRU stack for LRU replacement), allowing the OS to avoid replacement of recently-accessed MigrantStore pages. Due to locality and reasonable migration frequency, the number of unique pages touched between two migrations is small enough that both a small buffer captures most of the benefits (e.g., 20 entries) and the runtime overhead of the software is hidden well under the migration DMAs. In case the buffer capacity is exceeded, then the oldest entry is overwritten to provide a truncated list of recently-accessed pages and the software uses this truncated list to update its data structures.

### 3.3 Migration Hysteresis

Recall from Section 1 that to reduce the bandwidth and energy overhead of ineffectual migrations, we propose migration hysteresis where a page is migrated only after some number of accesses have occurred to the page in PCM. By reducing the number of ineffectual migrations, our hysteresis also reduces the software overhead on system performance and energy. One may think that replacement policies like LRU have built-in hysteresis obviating our explicit hysteresis. However, our hysteresis controls placement instead of replacement. While LRU prevents poor replacements but does not prevent unnecessary placement where performance and energy penalty are paid well before replacements occur (our results show that hysteresis achieves lower energy even in the presence of LRU).

To count the accesses to a page, we reuse the sub-block dirty bits (Section 3.4) which are not used by the sub-blocking scheme when the page is in PCM which is when our hysteresis needs to count. Blindly counting all accesses including cache hits may confuse our hysteresis because cache hits do not access the PCM and inject noise into the counts. Consequently, we count off-chip misses to a page in the hysteresis counter. Because current systems do not report off-chip misses to the TLB, we modify cache miss replay in the processor pipeline to enable our count. With the return of data to the pipeline for an off-chip cache miss, the off-chip miss is identified so that when the pipeline replays the access, the count is incremented in the TLB (piggy-backed with the update of the per-page reference bit as done in conventional TLBs upon every access, so there are no extra TLB accesses). The count is propagated to the page table upon a TLB eviction as done in current systems, incurring little overhead. A page is migrated when the count reaches the *hysteresis threshold*.

To ensure that migrations do not prevent non-migration, demand accesses (that are under the threshold) by swamping the PCM banks and memory bus, we give priority to non-migration accesses over migrations. We allow non-migration accesses to proceed in between consecutive cache-block-size bursts of a migration (Section 3.1). This prioritization balances the migrations' bandwidth and the non-migration accesses' delay.

One may think that migration hysteresis can also be done in the hardware cache, but there is a subtle issue. Migration hysteresis in the hardware cache would require tracking the access counts for all the blocks in physical memory, most of which are *not* in the cache. This tracking would add significant state overhead and complexity to the hardware cache. By piggybacking on the page tables, MigrantStore incurs much less overhead.

### 3.4 Page Sub-blocking

To save write energy and wear, we expose PCM's selective updates (Section 2.2) to the OS via *page sub-blocking* so that only the dirty sub-blocks within a page (e.g., 512-byte sub-blocks) are written back to PCM upon page eviction from MigrantStore. While the hardware cache performs this sub-blocking entirely in hardware, we expose this key aspect of PCM to the OS.

To implement sub-blocking, the page tables expand their per-page dirty to per-sub-block dirty bits, which are held in the TLBs as well. The sub-block dirty bits are significant only when the page is in MigrantStore and the bits are cleared when page is migrated into MigrantStore. Upon a write to a page in MigrantStore, the sub-block dirty bits are updated in the TLB (piggybacked with the update of the per-page dirty bit as done in conventional TLBs, so there is no extra TLB accses). Upon replacing a page from MigrantStore, the OS looks up the TLB and/or page table to determine which of the page's sub-blocks need to be written back to the PCM.

**Table 1: Hardware parameters**

| | |
|---|---|
| **Cores** | 8, in-order |
| **L1 Caches** | Split I&D, Private, 32K 4-way set associative, 64B cache block, 3 cycle hit, LRU |
| **L2 Cache** | Unified, Shared, Inclusive, 8M 8-way set associative, 8 banks, 37 cycle hit, LRU |
| **Coherence** | MESI Directory, Full bit vector in the L2 |
| **Memory Subsystem** | **Total latency = request/response queuing at controller + device latency + transfer latency** <br> **1 memory cycle = 10 CPU cycles** |
| **Bus** | 256 bits (total), 1 memory cycle |
| **SIMULATED SYSTEMS' LATENCIES and ENERGIES** | |
| **Base DRAM (DRAM-ideal)** | 8 GB, 22 memory cycles (reads and writes), total 64 banks, 64-byte interleaving, 32-entry bank queues |
| | 33 nJ row miss read/write; 16 nJ row hit read/write; 64 mW leakage |
| **Base PCM (PCM-only)** | Per-cell latency w.r.t. DRAM : 4x (reads), 12x (writes) |
| | 8 GB, 55 memory cycles (reads) and 143 cycles (writes), 64 banks, 64-byte interleaving, 32-entry bank queues |
| | Per-cell energy w.r.t. DRAM: 2x (reads), 43x (writes) |
| | 33 nJ row miss read; 36 nJ and 170 nJ row miss 64-byte cache block write and 512-byte sub-block write; 16 nJ row hit read/write; 6.4 mW leakage |
| **MigrantStore** | 128 MB DRAM, 16 memory cycles (reads and writes), total 16 banks, 64-byte interleaving, 32-entry bank queues |
| | 15 nJ row miss read/write, 4 nJ row hit read/write, 8 mW leakage |
| | negligible energy (0.025 nJ) for 20-entry *RAPid buffer* |
| **Hardware cache parallel (sequential)** | 128 MB DRAM, 16-way associative, 19 (25) memory cycles (reads and writes), total 16 banks, 64-byte interleaving, 32-entry bank queues |
| | 29 (15) nJ row miss read/write; 8 (4) nJ row hit read/write; 8 mW leakage |

Smaller sub-block size reduces the write energy and wear but also increase the space overhead of the extra dirty bits in the page tables. Because multiple adjacent cache blocks are often clean or dirty together due to spatial locality, the sub-blocks comprise many cache blocks which reduces the space overhead (e.g., 512-byte sub-blocks need 16 extra bits in the page table entry). As discussed in Section 3.3, because sub-block dirty bits are needed only for MigrantStore pages and hysteresis count only for PCM pages, we can reuse the same field in the page tables and TLB for the dirty bits and the counts. However, the sub-block dirty bits (e.g., 16) being more than the count (e.g., 4-6) may impose some space overhead on the PCM page tables which are significantly larger than the MigrantStore page tables. One option would be to employ separate page tables for the PCM and MigrantStore so that the PCM page tables hold the hysteresis count and the MigrantStore page tables hold the sub-block bits. The TLB would cache both the page tables and the TLB entries would still reuse the same field for the sub-block bits and the count.

Finally, because only the dirty sub-blocks of an evicted page are written back from MigrantStore to PCM, the page's clean sub-blocks must be intact in the PCM. However, under normal page migration from a source to destination, the migrating page fully vacates its space resources in the source so the space can be re-allocated to another page. To avoid losing the clean sub-blocks due to such re-allocation, we do not re-allocate the migrating page's space in PCM. This lack of re-allocation implies that for every page in the MigrantStore there is a stale page in the PCM. Because MigrantStore is much smaller than the PCM, this duplication overhead is small. The OS holds the stale pages in a data structure isolated from the page tables (i.e., the stale pages are not pointed to by any page table entry), so that upon an eviction from MigrantStore, the OS directs writeback from MigrantStore to the PCM to go to the appropriate stale page. The OS then changes the page table mapping to point to the updated stale page, turning the stale page into a current page.

## 4 Methodology

We simulate MigrantStore using Wisconsin GEMS-2.1 [8] built on top of Simics, a full-system simulator. We simulate a SPARC-based multicore running Solaris 10. For comparison, we also simulate the hardware cache in [12], the row buffers in [7], and page copying at the end of OS quanta [18]. The hardware parameters are given in Table 1. We obtain the PCM and DRAM latencies by combining information from PCM technology papers [13][2][6][3][17][4], CACTI's DRAM models [16] for the array decode, row buffers, and wiring latencies (these components are similar in PCM and DRAM), and Micron Sys-tem Power Calculator [9]. In Table 1, the total latencies are better than per-cell latencies, as described in Section 2.1. We account for latencies, bank and bus occupancies, and queuing at the controllers in all the memory components. Because PCM occupancies are large, we model an aggressively-banked base PCM-only system (many banks and deep queues) so that we do not unduly penalize the PCM-only case.

**Table 2: workloads**

| | |
|---|---|
| Commercial | **Apache:** 20k files (~**500MB**), 3200 clients, each with 25ms think time, warm up for ~1,500,000 transactions, 600 transactions executed. |
| | **Online Transaction Processing (OLTP):** PostgreSQL 8.3.7 database server, **5 GB** database with 25k warehouses, 128 users with 0 think time, and warm up the database for ~100k transactions before taking measurements for 200 transactions. |
| | **SPECjbb2000**: SPEC server workload v1.07, used Sun J2SE v1.5.0 JVM, simulated 1.5 warehouses/CPU with 0 think time, warm up for 350,000 transactions and measured for 10,000 transactions. (**300 MB**) |
| Scientific | **FFT**: Transpose computation in Fourier transform of $2^{24}$ complex data points (**256 MB**). |
| | **LU** decomposes a 4096 x 4096 matrix (between two barriers in one iteration of the main loop) (**128 MB**). |

We simulate the extra PCM and DRAM accesses needed for the page migrations in MigrantStore and the cache fills in the hardware cache. We carefully model open-page mode in both the PCM and DRAM for the page migrations and cache fills which perform sequential accesses and hence significantly benefit from open-page mode's row hits. Specifically, a migration in MigrantStore or a cache fill in the hardware cache involves reading the demand page from the PCM, reading the victim page from the DRAM, writing the demand page to the DRAM, and writing the victim page (if dirty) to the PCM. The first three operations exploit (1) parallel accesses across as many banks as needed to retrieve a page (all the 64 64-byte banks in PCM and all the 16 64-byte banks in DRAM) and (2) row locality for multiple accesses to the same bank (2 accesses in PCM and 8 accesses in DRAM). The last operation exploits selective updates to write back only the dirty sub-blocks. To ensure that our comparison with the hardware cache remains conservative, MigrantStore stalls the migration-triggering L2 miss for all the four operations to finish (otherwise the page tables would point to stale pages) whereas the hardware cache stalls the L2 miss only for the first operation assuming the rest of the operations are hidden in hardware. We tried returning the requested word(s) first for the hardware cache but the optimization is not effective due to stalls caused by further, immediate L2 misses to the DRAM cache block being transferred. Unlike latencies of some of the operations, *all* the operations incur bank occupancies in both MigrantStore and the hardware cache. As mentioned in Section 3.3, we trigger 64-byte bursts for the migrations and cache fills while allowing non-migration, demand accesses to proceed between the bursts. Each such burst looks up the L2 cache to invalidate or write back matching cache blocks. Overall, each migration incurs about 6000 cycles in the memory system. For MigrantStore, we charge 5000 cycles extra per migration for software overhead *in addition* to the memory system overhead. This overhead covers a fast trap startup (a few tens of cycles), initiating the DMAs for the migrations (tens of insructions), and updating the page table entries and shooting down TLB entries for the migrated pages (tens of instructions). The trap handler also scans the RAPid buffer and updates the LRU stack. Our code to scan a 20-entry RAPid buffer and update an LRU stack is about 60 static instructions and about 400 dynamic instructions (overall total of about 600 counting the other tens of instructions). Because OS trap code exhibits poorer locality than applications, we conservatively assume that the 600 instructions take 5000 cycles. One could hide the 400 instructions under the 6000 cycles of migration, but we conservatively assume no hiding.

We model PCM and DRAM energies also using the above sources. In Table 1, the total energies are better than per-cell energies, as described in Section 2.2. We assume that the PCM reads an entire row for non-migration reads such as L3 misses in *PCM-base* which does not use any DRAM (Section 5.1.1) and L2 misses under the hysteresis threshold with MigrantStore, migrations in MigrantStore and cache fills in the hardware cache; and that the PCM writes only cache blocks (and not entire rows) for non-migration writes (L2 writebacks in PCM systems without a DRAM cache and L2 writebacks that miss in MigrantStore due to hysteresis) and sub-blocks for dirty-page writebacks from MigrantStore or the hardware cache. For page migrations and cache fills, we carefully model row hits in both PCM and DRAM which avoid the large array access energies. Overall, each migration incurs about 8000 nJ in the memory system. For MigrantStore, we charge a software energy overhead of 3000 nJ per migration assuming 5 nJ for each of about 600 dynamic instructions in the trap software (Intel Xeon consumes 5 nJ/instruction). We note that because the software overhead of 5000 cycles and 3000 nJ are similar to the average per-migration memory system overhead of about 6000 cycles and 8000 nJ, reducing the software overhead via hysteresis is important (while these are averages, our simulations account for the actual per-migration overhead).

We use commercial and scientific workloads briefly described in Table 2. To account for statistical variations, we use enough randomly-perturbed runs to achieve 95% confidence [1].

# 5 Experimental Results

We first compare MigrantStore with other alternatives in terms of performance and energy. Then, we analyze the effectiveness of migration hysteresis and sub-blocking by comparing MigrantStore with and without these mechanisms. Next, we study the sensitivity of MigrantStore to its DRAM size. Finally, we compare MigrantStore with other alternatives in terms of wear.

## 5.1 Performance and Energy

We show commercial workloads in Section 5.1.1 and scientific workloads in Section 5.1.2.

### 5.1.1 Commercial workloads

In Figure 3, we compare the performance of plain PCM (*PCM-only*), plain DRAM (*DRAM-ideal*), PCM with an SRAM-based L3 cache (*PCM-base*), the hardware cache [12], enhanced row buffers (*Row-buffers*) [7], and page copying at the end of OS quanta (*OS-quanta-copy*) [18] against that of MigrantStore. We show commercial workloads here and scientific workloads later. The Y-axis shows performance normalized to that of *PCM-base* (higher is better). While *PCM-only* does not have any extra cache like the DRAM in MigrantStore and the hardware cache, *PCM-base* uses an area-equivalent SRAM-based, 24-MB L3 cache with 1-KB blocks. *DRAM-ideal* shows conventional DRAM if DRAM scaling were to continue. We show two variants of the hardware cache: one with sequential tag-followed-by-data access (*H/w-cache-seq*) employed in large caches to reduce energy by accessing only the matching way at the cost of extra latency [16] and the other with parallel tag-data access (*H/w-cache-par*) to optimize latency at the cost of extra energy to access all the set-associative ways in parallel. Both the hardware cache variants use a 128-MB DRAM, and are 16-way associative with page-size (8-KB) blocks, 512-byte sub-blocks, and LRU replacement. While the authors in [12] assume 1-GB DRAM cache for 32-GB PCM, we assume similar proportions but smaller sizes of 128-MB DRAM cache for 8-GB PCM to keep GEMS' cache warm-up from blowing up. The rest of the parameters are similar to those in [12]. Despite the size differences, our speedups for the hardware cache are similar to those in [12]. For the enhanced row buffers (*Row-buffers*), we assume 8, 2048-byte-wide row buffers per PCM bank (the energy-performance optimal point in [7] is 4, 512-byte-wide row buffers per bank). *MigrantStore* also uses a 128-MB DRAM with 8-KB pages, 512-byte sub-blocks and hysteresis threshold of 16. We give the latencies and energies of the various schemes in Table 1. *OS-quanta-copy*, the last scheme, also uses a 128-MB DRAM. Though larger DRAMs are available today, we choose 128 MB DRAM for the hardware cache and MigrantStore to emulate the future where data will increase but DRAM will have stopped scaling.

Due to the lack of an L3 cache, *PCM-only* performs worse than *PCM-base*. *DRAM-ideal* performs better than *PCM-base* because PCM has longer latencies and higher occupancies than DRAM and *PCM-base*'s L3 cache is not large enough. Because DRAM is expected to stop scaling soon, this comparison is only to establish the opportunity for the rest of the schemes.

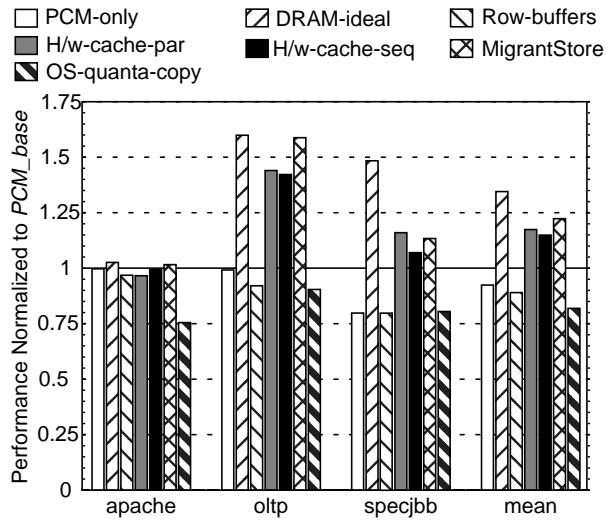

**FIGURE 3: Performance of commercial workloads**

*Row-buffer* performs worse than *PCM-base* due to (1) high row miss rate (75-96%) and (2) the fact that the L2 misses that incur a row miss need to wait for not only the PCM-array read latency but also the long PCM-array write latency of the row writeback if dirty (see the footnote in Section 2.2). To be certain, we validated our *row-buffer* implementation by running *ocean* and *radix* from the SPLASH suite where *row-buffer* achieves 2.77 and 1.09 speedups over *PCM-only* (6% and 42% row miss rate), respectively, which are in line with the *row-buffer* paper. However, our commercial workloads exhibit much poorer row locality than SPLASH.

Our multitheaded workloads mostly absorb the modestly longer latency of *H/w-cache-seq* as compared to *H/w-cache-par*, resulting in the variants achieving similar performance. Both variants perform better than *PCM-base* due to their DRAM caches. *MigrantStore* performs 7% better than *H/w-cache-seq*. Overall, *MigrantStore*'s larger DRAM achieves about 24% better performance over *PCM-base*'s SRAM-based L3. Finally, *OS-quanta-copy* performs worse than *PCM-base* because copying frequently-written pages into DRAM at the end of the quanta is too late by when many accesses have incurred PCM latencies and copying frequently-written pages ignores frequently-read pages. (Section 2.4).

In Figure 4, we compare the energy of the same schemes. The Y-axis shows energy normalized to that of *PCM-base* (lower is better). We break down energy into leakage, dynamic and software overhead (for MigrantStore, as discussed in Section 4). PCM's leakage is much lower than DRAM's (Table 1). The open-page mode accesses for MigrantStore's migrations and the hardware cache's cache fills (Section 4) improve both performance and energy by minimizing the high-energy row retrievals from the (DRAM or PCM) arrays.

Because *PCM-base*'s SRAM-based L3 cache consumes high energy, *PCM-only, DRAM-ideal* and *row-buffers* con-

**Table 3: Detailed metrics** (cycle* = execution cycle in *PCM-base*)

| Bench-marks | DRAM-ideal | PCM-only | PCM-base | H/w-cache-seq | | MigrantStore | | | NoH-noS | | H-S128 |
|---|---|---|---|---|---|---|---|---|---|---|---|
| | #busy bank/cycle* | #busy bank/cycle* | #busy bank/cycle* | #busy bank/cycle* | % miss rate | #busy bank/cycle* | % miss rate | %migrations/L2 miss | #busy bank/cycle* | % miss rate | %writeback reduction |
| apache | 0.53 | 3.61 | 0.5 | 0.4 | 1.03 | 0.39 | 6.11 | 0.31 | 0.44 | 0.80 | 44.48 |
| OLTP | 0.7 | 3.91 | 0.6 | 0.54 | 1.39 | 0.45 | 5.83 | 0.44 | 0.5 | 0.78 | 20 |
| specjbb | 0.91 | 5.74 | 0.9 | 0.8 | 2.9 | 0.64 | 16.61 | 1.02 | 0.87 | 2.98 | 8.85 |

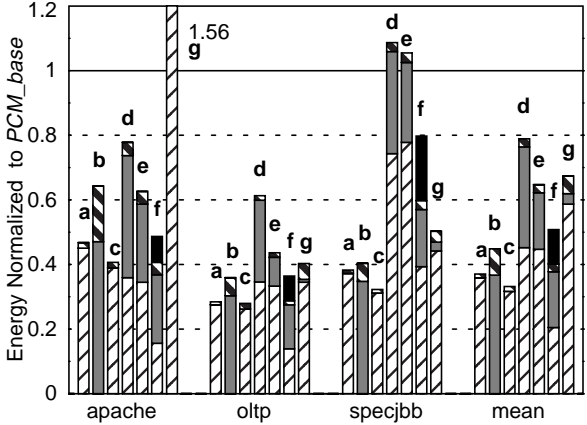

a - PCM-only
b - DRAM-ideal
c - Row-buffers
d - H/w-cache-par
e - H/w-cache-seq
f - MiigrantStore
g - OS-quanta-copy
■ S/w overhead
▨ Leakage
▨ DRAM-dynamic
▨ PCM-dynamic

**FIGURE 4: Energy of comercial workloads**

sume lower energy than *PCM-base*. While *H/w-cache-seq* consumes lower energy than *H/w-cache-par* by accessing only the matching way, *MigrantStore* consumes 25% and 56% less energy than *H/w-cache-seq* and *PCM-base*, respectively. *H/w-cache-par* incurs associative accesses (extra DRAM energy) and ineffectual cache fills (extra DRAM and PCM energy). *H/w-cache-seq* does not incur associative accesses but incurs ineffectual cache fills. Because there is no associative access, the per-access energy is equal to MigrantStore's with a negligible tag overhead (Table 1). MigrantStore does not have associative accesses and reduces ineffectual migrations via hysteresis. Because *OS-quanta-copy* copies much fewer pages than MigrantStore and ends up mostly accessing the PCM instead of the DRAM, *OS-quanta-copy*'s energy consumption is similar to that of *PCM-only* (as is performance in Figure 3). We do not charge any software overhead for the software copying in *OS-quanta-copy*. In *apache*, most of the copying is too late be useful but consumes energy.

We explain MigrantStore's energy using Table 3 which shows the average number of busy banks in *PCM-base*, *DRAM-ideal*, *H/w-cache-seq* and *MigrantStore* computed by averaging over *PCM-base*'s execution cycles. Because the execution times of these systems are quite different (Figure 3), averaging over each system's execution time would distort the true counts of the systems' busy banks. The table shows the fraction of L2 misses that further miss in *H/w-cache-seq* and in *MigrantStore (%miss rate)*. In *H/w-cache-seq,* all the cache misses trigger a cache fill (i.e.,

*%miss rate* = cache fills per L2 miss) and in *MigrantStore* only those MigrantStore misses that are above the hysteresis threshold trigger a migration. We also show the fraction of L2 misses that trigger migrations in *MigrantStore (%migrations/L2 miss)*. We discuss the last two columns (*NoH-noS* and *H-S128*) in Section 5.2. The numbers for *H/w-cache-par* are similar to those of *H/w-cache-seq*, as expected, and hence not shown.

*PCM-only* has many more busy banks than *DRAM-ideal*. Given that there are 64 banks, the higher number of busy banks in *PCM-only* confirms the bandwidth pressure on *PCM-only*. *PCM-base*, *H/w-cache-seq,* and *MigrantStore* considerably reduce this pressure via caching. Because MigrantStore does not migrate on every MigrantStore miss whereas *H/w-cache-seq* does, *H/w-cache-seq* achieves lower cache miss rate than *MigrantStore (%miss rate* in Table 3). However, many of the fills are ineffectual and are avoided by MigrantStore's hysteresis, as confirmed by the fact that despite the higher miss rates, MigrantStore performs better than *H/w-cache-seq* (Figure 3). *H/w-cache-seq*'s fills (fills per L2 miss = *%miss rate* in Table 3) are more than MigrantStore's migrations (%*migrations/L2 miss* in Table 3). These fills consume both PCM and DRAM energy, resulting in *H/w-cache-seq*'s higher energy than MigrantStore.

In summary, we have shown that MigrantStore performs better than all the compared alternative schemes and consumes less energy than all the DRAM-based alternatives. Further, recall that MigrantStore avoids the SRAM tag cost overhead of the hardware cache.

### 5.1.2 Scientific workloads

In Figure 5, we compare the performance and energy of the same schemes as before running scientific benchmarks, SPLASH's FFT and LU. The Y-axis shows both performance and energy normalized to those of *PCM_base*. Because of these benchmarks' much lower memory pressure than commercial workloads, there is little difference in the various systems' performance. Due to good row locality, *row-buffers* perform 16% better than the rest in *lu*. In terms of energy, *PCM-only*, *row-buffers*, *h/w-cache-seq*, *h/w-cache-par*, and *MigrantStore* perform better than *PCM-base*, for the same reasons as in the commercial workloads. Due to low L2 miss rates in *lu*, the dynamic energy is much lower than the DRAM leakage in *DRAM-ideal*, *h/w-cache-seq*, *h/w-cache-par*, and *MigrantStore* (*DRAM-ideal*'s leakage is pronounced due to its large DRAM).

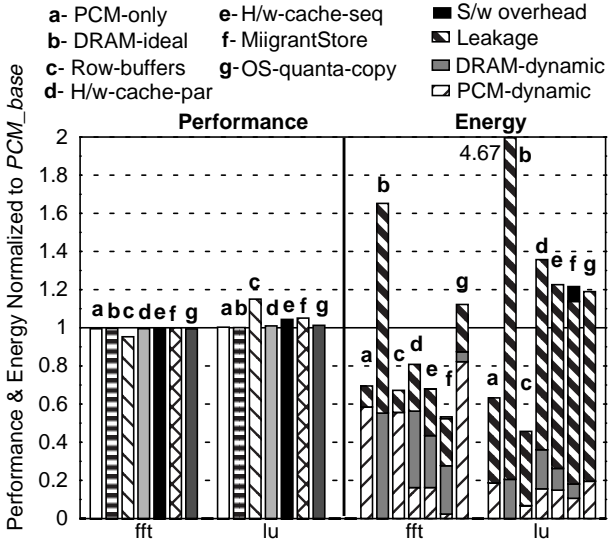

**FIGURE 5: Performance & energy of scientific workloads**

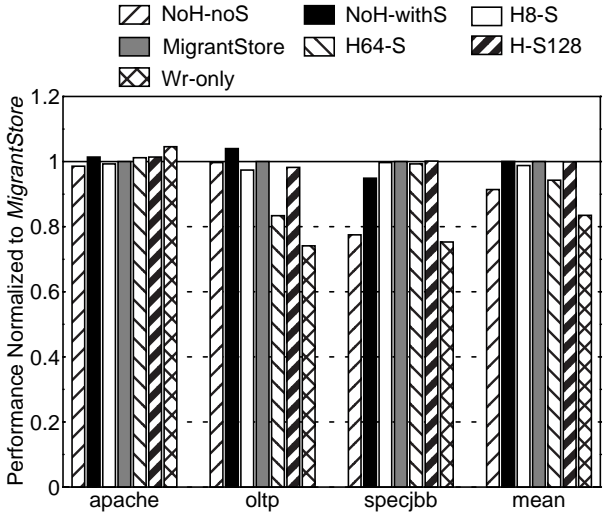

**FIGURE 6: Performance impact of hysteresis & sub-blocking**

Because the performance of the systems are similar for the scientific workloads, we do not analyze these workloads any further.

### 5.2 Hysteresis, sub-blocking & RAPid buffer

In Figure 6, we compare the performance of many variants of MigrantStore: without hysteresis and without sub-blocking (*NoH-noS*), without hysteresis and with 512-byte sub-blocking (*NoH-withS*), with hysteresis threshold and sub-block size in bytes of (1) 16 and 512 (default, *Migrant-Store*), (2) 8 and 512 (*H8-S*), (3) 64 and 512 (*H64-S*), and (4) 16 and 128 (*H-S128*), and migration only on writes with hysteresis threshold of 16 and 512-byte sub-blocking (*Wr-only*). The Y-axis shows the performance of these variants normalized to that of the default. In Figure 7, we show the variants' energy.

Both *NoH-noS* and *NoH-withS* perform variedly across our workloads (Figure 6) and more importantly incur significant energy degradations (Figure 7) compared to *MigrantStore* (the default), highlighting the importance of hysteresis. The no-hysteresis configurations migrate on all DRAM misses like the hardware cache and incur many more migrations than the default *MigrantStore (NoH-noS*'s migrations per L2 miss = *%miss rate* in Table 3). These migrations are fewer than those of *H/w-cache-seq*'s cache fills (*H/w-cache-seq*'s fills per L2 miss = *%miss rate* in Table 3) due to better miss rates of the fully-associative MigrantStore versus the 16-way-associative hardware cache. In the absence of sub-blocking, *NoH-noS*'s PCM energy component is large due to full-page PCM writes. *H8-S* and *MigrantStore* bars show that hysteresis thresholds of 8 and 16 perform well (Figure 6), confirming that the hysteresis is fairly stable across the 8-16 range of threshold values. *H64-S* incurs performance loss (Figure 6) due to considerably delayed migrations.

*H-S128* compares our coarse 512-byte sub-blocks against the fine 128-byte sub-blocks and shows that the finer granularity does not considerably change performance (Figure 6) but improves energy (Figure 7). This improvement is explained by the dirty-page writeback reduction in *H-S128* as a percent of all PCM traffic in the default *MigrantStore* using 512-byte sub-blocks (*H-S128*'s *%writeback reduction* in Table 3), and exposes the trade-off between page-table space overhead and the writeback traffic. Finally, we see that migrating only on writes (*Wr-only*) improves energy due to fewer migrations but incurs considerable performance loss due to not migrating read-only or read-mostly pages (also a problem for *OS-quanta-copy* in Figure 3).

We note that MigrantStore's 25% energy advantage over *h/w-cache-seq* (Figure 4) would disappear without hysteresis and sub-blocking as seen by the nearly 120% energy overhead of *NoH-noS* over MigrantStore (Figure 7). This comparison highlights the importance of hysteresis and sub-blocking.

In Figure 8, we isolate the performance and energy impact of MigrantStore replacement policies by comparing

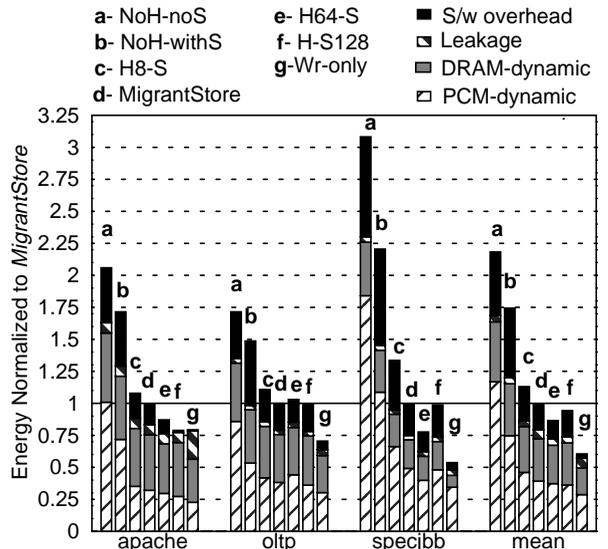

**FIGURE 7: Energy impact of hysteresis & sub-blocking**

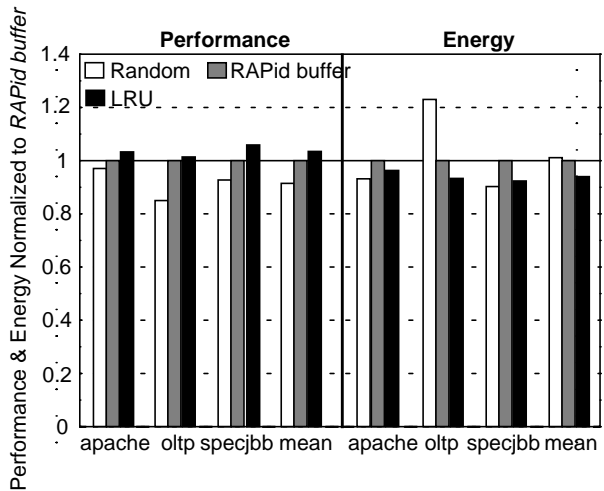

FIGURE 8: Performance & energy impact of MigrantStore replacement policiies

MigrantStore with perfect LRU, 20-entry RAPid buffer (which approximates perfect LRU whenever the buffer's capacity is exceeded), and random replacement. The Y-axis shows the performance and energy normalized to that of the default MigrantStore with the RAPid buffer. There are gaps in average performance and energy of about 11% and 7%, respectively, between *random* and *LRU*. A 20-entry RAPid buffer is sufficient to bridge this gap. The energy gap is narrower than the performance gap because the software overhead incurred by *LRU* to update its data structures is not present in *random*.

### 5.3 Wear

We evaluated MigrantStore's wear and found that MigrantStore's lifetime is similar to that of the hardware cache (see Appendix for details).

## 6 Conclusion

Architectural and system support will likely be required to exploit PCM's advantages (scalability and energy) and alleviate its disadvantages (performance and reliability). Previous DRAM-based hardware cache incurs significant bandwidth and energy overhead due to ineffectual caching where blocks are evicted before sufficient number of accesses. Bandwidth and energy are two key but scarce resources in modern multicores. We employed intelligent OS policies to avoid the hardware cache's ineffectual cache fills. We proposed a DRAM-PCM hybrid memory architecture that leverages virtual memory so that the OS migrates pages on demand from the PCM to DRAM, which is called the *MigrantStore*. Conventional OS placement and replacement policies for managing physical memory cause MigrantStore to incur significant performance and energy degradations. We proposed two ideas to address these issues. First, to reduce the energy, bandwidth, and wear overhead of ineffectual migrations, we proposed *migration hysteresis*. Second, to reduce the software overhead of good replacement policies, we proposed *recently-accessed-page-id (RAPid) buffer*, a hardware buffer to track the addresses of recently-accessed MigrantStore pages.

Using simulations of commercial workloads, we showed that MigrantStore performs better than all the compared alternative schemes and consumes less energy than the DRAM-based schemes (e.g., a 128-MB MigrantStore improves energy and performance by 25% and 7%, respectively, while achieving similar wear as compared to a 128-MB hardware cache). Being a part of physical memory, MigrantStore avoids the SRAM tag cost overhead of the hardware cache. Our experiments showed that migration hysteresis and page sub-blocking are crucial for MigrantStore's performance and energy. MigrantStore's performance, energy, wear, and cost make it an attractive option for architecting future memory systems using PCM technology. By leveraging virtual memory, MigrantStore also opens up the possibility for interesting architecture-operating systems synergies to be exploited in PCM-based systems.

## Appendix

## Sensitivity to DRAM size

In Figure 9, we vary MigrantStore's size (i.e., DRAM size) as 128 MB and 256 MB. The Y-axis shows the mean performance and energy of *apache*, *oltp*, and *specjbb* normalized to that of *PCM-base*. As the DRAM size increases, *MigrantStore* improves performance, as expected. The larger DRAM achieves lower miss rate than the smaller DRAM so that fewer accesses go to the PCM, resulting in lower energy for the larger DRAM.

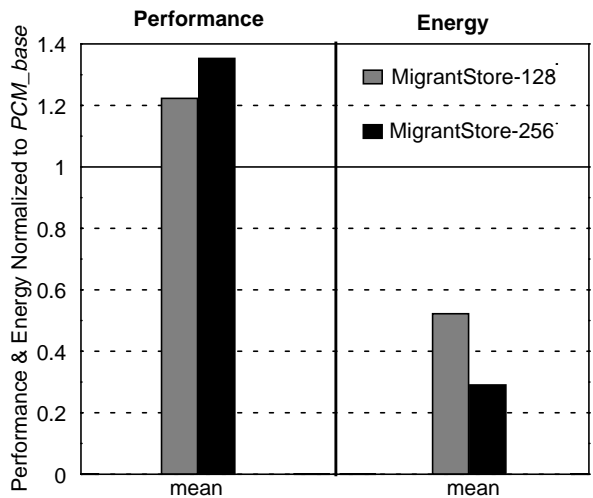

**FIGURE 9: Sensitivity to DRAM size**

## Wear

In Figure 10, we show the distribution of writes per block for *PCM-only*, *PCM-base*, *H/w-cache-seq* (128MB), and *MigrantStore* (128MB), as five graphs, one for each of our workloads. The X-axis shows the number of writes to a given cache block in PCM and the Y-axis shows the cumulative percent of all blocks accessed in the simulation run. Because we did not observe any systematically biased writes to each page's block#0, as reported by the hardware cache paper (Section 2.4), we do not include the hardware cache's block rotation scheme for *h/w-cache-seq*. *PCM-base*, *h/w-cache-seq* and *MigrantStore* achieve much fewer number of per-block writes than *PCM-only*. By avoiding ineffectual migrations, MigrantStore incurs fewer dirty-page evictions from the DRAM, and thereby achieves fewer per-block writes than *h/w-cache-seq*. In the case of MigrantStore, the extra writes to the PCM due to L2 writebacks to non-migrated pages spread out to different memory blocks and hence avoid worsening the per-block writes to the PCM. This spreading occurs due to good L1 and L2 caching which prevents repeated writes to a given memory block from reaching the PCM. With the scientific workloads, *fft* and *lu*, all the systems incur fewer than one write to a large fraction of the memory blocks due to the workloads' low memory pressure.

While the graphs show the number of per-block writes, lifetimes, in general, depend upon both the rate at which the workloads write to the PCM and the distribution of writes per block. Taking write rates into account, Table 4 summarizes the *worst-case lifetime* of 99.99% of memory blocks for the three schemes running the different workloads. We use 99.99% of the blocks to remove any pathological cases that hurt the base case much more than the other systems. We assume that random OS page churn smooths out writes across physical frames so that the number of writes to a frame does not exceed the maximum among 99.99% of frames. Because real systems experience significant amount of page churn, this assumption is likely to be conservative. Further, if random page churn does not address the 0.01% then an active PCM-to-PCM migration can do so.

We define worst-case lifetime of a set of blocks as the lifetime of the subset of blocks that are written the most number of times. Table 4 shows the maximum number of writes to any of the 99.99% of blocks during the simulation run and the expected normalized lifetime assuming that PCM can endure $10^9$ writes. To normalize the write rates across these systems which vary in speed, we use *PCM-base*'s execution time for all the systems, so that the rates are not biased by execution speed. Compared to *PCM-only*, *PCM-base*, *H/w-cache-seq* and *MigrantStore* significantly reduce the maximum number of writes. MigrantStore further reduces dirty-page evictions due to ineffectual migrations to achieve even fewer writes. *fft* and *lu* have longer lifetimes than the commercial workloads because *fft* and *lu* exert significantly lower memory pressure, and hence have lower write rates.

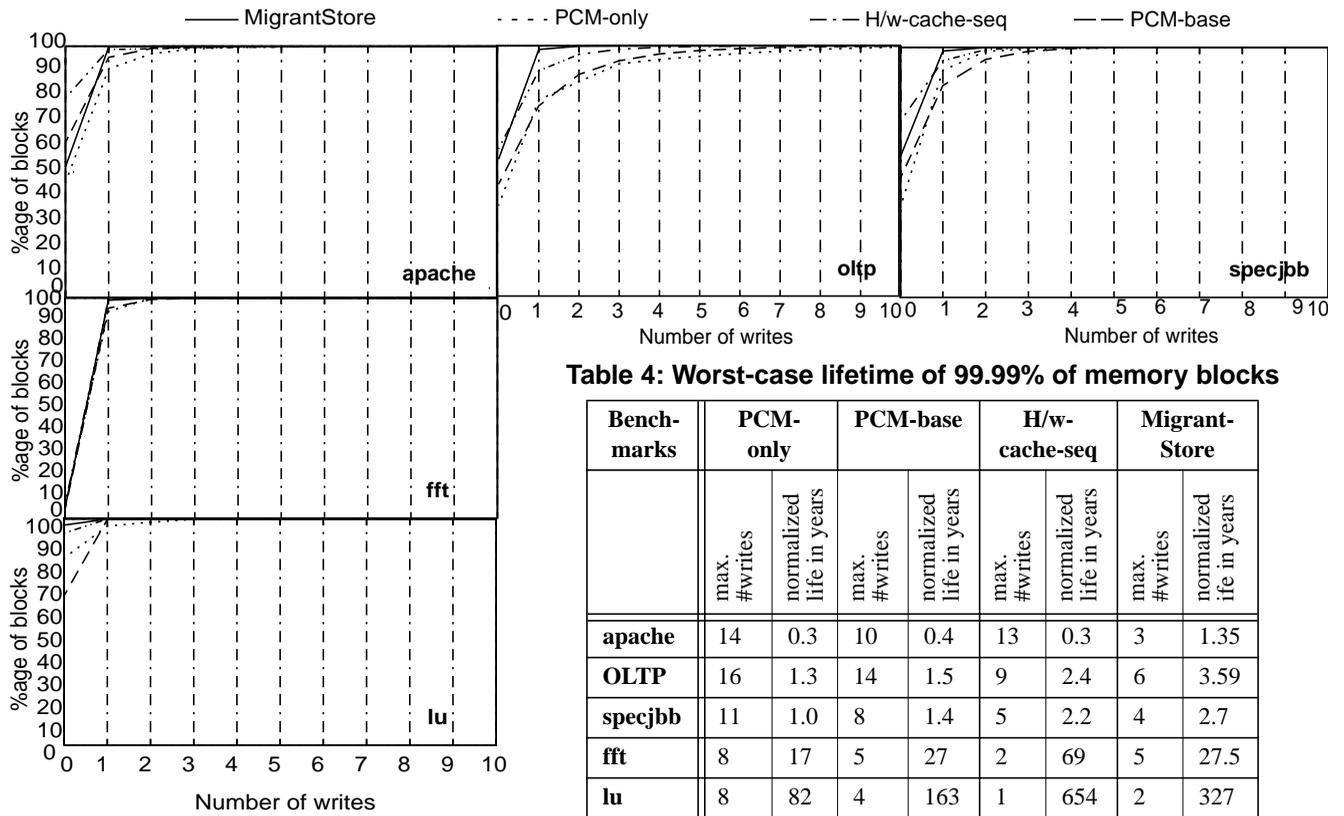

**FIGURE 10: Wear**

**Table 4: Worst-case lifetime of 99.99% of memory blocks**

| Bench-marks | PCM-only | | PCM-base | | H/w-cache-seq | | Migrant-Store | |
|---|---|---|---|---|---|---|---|---|
| | max. #writes | normalized life in years | max. #writes | normalized life in years | max. #writes | normalized life in years | max. #writes | normalized life in years |
| apache | 14 | 0.3 | 10 | 0.4 | 13 | 0.3 | 3 | 1.35 |
| OLTP | 16 | 1.3 | 14 | 1.5 | 9 | 2.4 | 6 | 3.59 |
| specjbb | 11 | 1.0 | 8 | 1.4 | 5 | 2.2 | 4 | 2.7 |
| fft | 8 | 17 | 5 | 27 | 2 | 69 | 5 | 27.5 |
| lu | 8 | 82 | 4 | 163 | 1 | 654 | 2 | 327 |